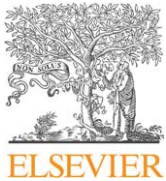

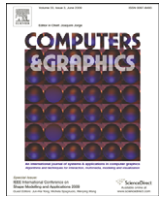

Technical Section

# Real-time visio-haptic interaction with static soft tissue models having geometric and material nonlinearity

Igor Peterlík [a,*], Mert Sedef [b], Cagatay Basdogan [b], Luděk Matyska [a]

[a] Faculty of Informatics, Masaryk University, Brno, Czech Republic
[b] College of Engineering, Koc University, Sariyer, Istanbul, Turkey

### ABSTRACT

Realistic soft tissue models running in real-time are required for the development of computer-based surgical training systems. To construct a realistic soft tissue model, finite element (FE) modeling techniques are preferred over the particle-based techniques since the material properties can be integrated directly into the FE model to provide more accurate visual and haptic feedback to a user during the simulations. However, running even a static (time-independent) nonlinear FE model in real-time is a highly challenging task because the resulting stiffness matrix (K) is not constant and varies with the depth of penetration into the model.

We propose a new computational approach allowing visio-haptic interaction with an FE model of a human liver having both nonlinear geometric and material properties. Our computational approach consists of two main steps: a pre-computation of the configuration space of all deformation configurations of the model, followed by the interpolation of the precomputed data for the calculation of the nodal displacements and reaction forces that are displayed to the user during the real-time interactions through a visual display and a haptic device, respectively. For the implementation of the proposed approach, no a priori assumptions or modeling simplifications about the mathematical complexity of the underlying soft tissue model, size and irregularity of the FE mesh are necessary. Moreover, it turns out that the deformation and force responses of the liver in the simulations are heavily influenced by the selected simulation parameters, such as the material model, boundary conditions and loading path, but the stability of the visual and haptic rendering in our approach does not depend on these parameters. In addition to showing the stability of our approach, the length of the precomputations as well as the accuracy of the interpolation scheme are evaluated for different interpolation functions and configuration space densities.

© 2009 Elsevier Ltd. All rights reserved.

## 1. Introduction

Simulating real-time visual interactions with deformable objects in virtual environments is an attractive, but computationally demanding area of research with applications to medical simulation and training. Integrating haptics into these simulations imposes more stringent constraints on the speed and accuracy of computations. While the refresh rate needed for flawless visualization is about 30 Hz, the rate required for haptic rendering of deformable objects is close to 1 kHz due to the higher sensitivity of our haptic channel. Moreover, organ-force models used in medical simulations must be convincing and therefore based on real physics for positive training transfer. However, developing realistic organ-force models is a highly challenging task not only because of the nonlinearity, rate, and time dependence of an organ's material properties but also because of its layered and nonhomogeneous structure. The relations from the theory of elasticity are usually employed when establishing the mathematical formulation of the problem that is finally solved by some complex computational method such as finite elements (FE). To implement the FE method, the geometric model of the organ is divided into surface or volumetric elements, the properties of each element are formulated, and then the elements are combined to compute the organ's deformation states under the influence of external forces such as the ones applied by the surgical instruments. A major advantage of FE modeling is that it uses continuum mechanics and has a solid mathematical foundation. On the basis of the partial differential equations and the selected constitutive relation, FE models can accurately approximate static and dynamic deformations of an organ having linear and nonlinear material properties. Another advantage is that FE models require only a few material parameters to describe the realistic response of the organ. However, it is known that simulating "ideal" soft tissue behavior using FE approach within the haptic loop is far beyond the capabilities of today's computers. There have been several attempts to address this issue such as the simplification of the underlying mathematical models or employing precomputations before the real-time interaction occurs. Below, we provide a brief summary of the FE approaches used for simulating deformable behavior of soft objects. The readers can find more extensive coverage of the soft tissue modeling techniques applied to medical simulation in our recent survey paper [1].

* Corresponding author.
  E-mail addresses: peterlik@ics.muni.cz (I. Peterlík), lsedef@ku.edu.tr (M. Sedef), cbasdogan@ku.edu.tr (C. Basdogan), ludek@ics.muni.cz (L. Matyska).

### 1.1. Related work

One of the first studies in the field of soft-tissue modeling is reported in [2]. The technique utilizes the FE approach for implementation of a linear model. An update of 30 Hz, equivalent to the visual refresh rate is achieved using the condensation technique. A small-deformation model suitable for laparoscopic surgery is implemented in [3]. In [4] modal analysis is applied to a linear FE model to reduce the number of computations and achieve real-time visual and haptic update rates. A method based on small area of contact is proposed in [5] for real-time interaction with a deformable static FE model running at haptic refresh rate. While the complexity of on-line computations does not depend on the size of the mesh in this paradigm, it can only be implemented with a linear FE model. Extensive research on soft-tissue modeling is performed within Epidaure project [6]. First, a linear model employing displacement-driven haptic interaction is proposed. It is based on a static superposition of unit displacements which are precomputed in advance. Force extrapolation is used for haptic rendering to compensate for the low update rate of the model [7]. Further, to allow topological changes (such as tearing or suturing), a dynamic mass-tensor model is proposed to simulate linear but anisotropic material. The mass-tensor model is extended by including geometrical nonlinearities via Green–St. Venant strain tensor in [8], however, linear material properties are assumed and reasonable haptic update rates are achieved by utilizing the force extrapolation technique again. The point-associated finite-field (PAFF) approach presented in [9], also called the finite-spheres method, is a meshless FE approach applied to surgical simulation. PAFF supports simulation of large deformations as well as topology modifications such as cutting, but the brute-force implementation of this technique is still computationally intensive. A finite element model handling geometric nonlinearities is proposed and implemented in [10]. The model employs mass-lumping for real-time simulation of dynamic behavior. Both geometrical and material nonlinearities based on Mooney–Rivlin material are modeled in [11]. The implementation is based on dynamic progressive meshes which allow realistic modeling of the deformation in the vicinity of the contact. Although both types of nonlinearity are considered, the initial mesh utilized during the precomputation must be dense enough to cover all the expected deformations. A linear viscoelastic model is developed in [12]. The approach is also based on precomputations. First, the force response of each surface node and the displacement responses of its neighboring nodes are recorded for a unit step displacement applied to the surface node for 30 s. Second, the recovery displacement responses of each surface node and its neighboring nodes are recorded for a unit step force applied to the surface node for 10 ms. These two sets of data are then used in tandem for computation of nodal displacements and interaction forces during real-time simulations. Recently, a new approach based on precomputation is reported in [13]. Although this approach allows the interaction with complex body having nonlinear geometry properties, it is implemented with a linear material model (St. Venant) only. The same authors have also investigated 6DoF haptic rendering of contact interactions with a deformable FE model in [14]. An object with a complex 3D geometry is considered in the simulations, nevertheless, the proposed computational approach is suitable for one type of material model only.

To summarize, most of the approaches presented above either assume linear material and geometric properties or make modeling assumptions or simplifications to implement nonlinearities with FE models, but then the accuracy of solutions are jeopardized. Integrating both geometric and material nonlinearities into a static FE model and simulating realistic visual and haptic interactions with a deformable mesh having complex geometry at real-time rendering rates still remains an open problem.

### 1.2. Our contribution

If linear FE models are used, only small deformations are rendered realistically. However, it is known that soft organ tissues show large deformations under the influence of external forces and the small deformation assumption is not valid for modeling their behavior. In addition to geometric nonlinearities, material properties of soft organ tissues are also nonlinear and our sense of touch is sensitive to these nonlinearities as demonstrated experimentally in [15]. If modeling assumptions are made to solve nonlinear FE models in real-time, undesired artifacts may appear in the simulations depending on the level of accuracy of these assumptions. We have developed a new precomputation approach allowing haptic interaction with static FE models having both geometric and material nonlinearities. The approach is based on a notion of traveling through a configuration space which is precomputed in advance. In the past, we used this approach for the simulation of haptic interactions with biomolecules [16].

The main difference between our approach and the earlier precomputation approaches is that we do not make any assumptions about the mathematical complexity of the underlying soft tissue model, size and irregularity of the FE mesh and the computational complexity of the corresponding solution method. Therefore, also nonhomogeneity and anisotropic tissue models can be considered, as the approach being proposed is independent of the underlying properties of the FE model. On the other hand, the current implementation of the proposed approach has some limitations. First, only static model of the tissue is considered and no topology changes are currently allowed. Second, the approach is demonstrated using a single point interaction only. Finally, the force vector is interpolated during the real-time interaction, so an approximation error is introduced affecting the realistic force response of the model.

Having a formulation of the boundary problem derived within the theory of elasticity, the finite element method is applied in a standard manner resulting in large system of nonlinear algebraic equations. These equations must be solved iteratively, which is not possible to do in real-time. In our approach, all these expensive iterative computations are performed in advance during an off-line precomputation phase. This results in a large number of precomputed configurations which are stored to a data file. An advantage of this approach is that any convergence issues which are quite usual in the area of nonlinear modeling can be addressed within the off-line calculations. During the real-time interaction, the precomputed configurations are interpolated to calculate the displacements of the nodes and the reaction force vector displayed to the user through a haptic device. The interpolation approach used for the estimation of the force vector is simple enough to execute in the haptic update loop. It is important to emphasize that the proposed approach is not based on the superposition principle, which is not valid for the nonlinear models, however, the configurations are "sampled" by standard FE method.

The outline of the paper is the following. First, the mathematical background of the problem is briefly presented together with iterative solution method. Further, the integration of the material parameters of the soft tissues into the model is described. In the fourth section, the simulation parameters are identified and the technique based on the precomputation and reconstruction is presented.

The fifth section is dedicated to the implementation of the approach. The distributed version of the precomputation is shown and the overview of the computational experiments is given. This is followed by evaluation section where the behavior of the model is briefly studied, the length of the precomputation phase and the accuracy of the reconstruction are discussed.

## 2. Mathematical background and numerical solution

In this section, we briefly present the mathematical background of the deformation modeling of the soft tissues. First, the physical formulation of the problem is sketched and then, the numerical solution is described.

### 2.1. Physical representation of deformations

Since the realistic behavior of the model is one of the key requirements in surgical simulation, the relation between nodal displacements of deformable body and applied surface forces is derived based on the theory of elasticity. Denoting an arbitrary particle in a deformable body as $\mathbf{x}$, the deformation is defined as a mapping $\varphi : \Omega \rightarrow \Omega'$ from the undeformed to the deformed domain given as $\varphi(x_i) = x_i' = x_i + u_i$ where $u_i$ are components of displacement defined as a function $u_i(x_1, x_2, x_3)$.

The important entity which can be regarded as internal measure of the strain is represented by the Green–St. Venant strain tensor $\gamma_{ij}$ which is defined as

$$\gamma_{ij} = \tfrac{1}{2}(u_{i,j} + u_{j,i} + u_{m,i}u_{m,j}) \tag{1}$$

where the notation convention $u_{i,j} = \partial_j u_i = \partial u_i / \partial x_j$ for derivatives together with the summation convention are applied.

Throughout this paper, the static equilibrium between the external and internal forces in the deformed configuration $\Omega'$ is considered. The former usually constitutes the surface traction forces $\mathbf{g}'$ and applied body forces $\mathbf{f}'$, whereas the latter are calculated by Cauchy stress tensor $\tau'$. Based on the conservation laws it can be shown that the static equilibrium of the deformed domain $\Omega'$ in the Eulerian system is represented by

$$\mathbf{0} = \mathbf{f}'(\mathbf{x}') + \nabla \cdot \tau'(\mathbf{x}'), \quad \mathbf{x}' \in \Omega' \tag{2}$$

$$\mathbf{g}'(\mathbf{x}') = \mathbf{T}'(\mathbf{x}', \mathbf{n}'), \quad \mathbf{x}' \in \partial\Omega' \tag{3}$$

where $\partial\Omega'$ is boundary of the domain $\Omega'$, $\mathbf{n}'$ is surface normal in $\mathbf{x}'$ and $\mathbf{T}'$ is Cauchy stress vector related to the Cauchy stress tensor via $\mathbf{T}' = \tau'\mathbf{n}'$. The above equations can be transformed from the deformed coordinate system which is not a priori known to the undeformed coordinate system by Piola transformations [17]. Then, the equilibrium equation can be formulated using the symmetric second Piola stress tensor $S_{ij}$ as

$$-f_i = \partial_j\{(\delta_{im} + u_{i,m})S_{mj}\} \tag{4}$$

where the summation convention is applied. The connection between the strain tensor $\gamma_{ij}$ and the stress tensor $S_{ij}$ is provided by the constitutive law [18]. In this paper, we focus on hyperelastic materials which are typically used for the soft-tissue modeling. The hyperelastic materials are characterized by the concept of stored energy function $W$, a scalar function which fully determines the stress/strain relationship. The function $W$ is coupled with the second Piola stress tensor as

$$S_{ij} = \frac{\partial W}{\partial \gamma_{ij}} \tag{5}$$

There are multiple definitions of $W$ suitable for various types of materials [17]. In our study, we implement and compare the St. Venant–Kirchhoff material having Lamé coefficients $\lambda$ and $\mu$ defined in terms of the strain tensor $\gamma$ as

$$W = \frac{\lambda}{2}\gamma_{ii}\gamma_{jj} + \mu\gamma_{ij}\gamma_{ji} \tag{6}$$

and the Mooney–Rivlin material having two material constants $C_{10}$ and $C_{01}$

$$W = 2[\gamma_{ii}(C_{10} + 2C_{01}) + 2C_{01}[\gamma_{ii}\gamma_{jj} - \gamma_{ij}\gamma_{ji}]] \tag{7}$$

where the nonlinear incompressibility condition $\det(\delta_{ij} + \partial_j u_i) = 1$ must be applied.

In the case of St. Venant material combined with the large-strain tensor $\gamma_{ij}$, the resulting law is referred as being nonlinear in geometry and linear in material, whereas if the Mooney–Rivlin stored energy with the incompressibility conditions is used, the resulting law is regarded as nonlinear in geometry and material.

### 2.2. Finite element formulation and numerical solution

The governing system introduced in the previous section consists of the partial differential equation given by Eq. (4) which is to be solved over some domain $\Omega$ which can be irregular and complex. Such a problem is usually solved by finite element method which consists of several steps. First, a weak formulation of the problem is derived based on multiplication of both sides by weight functions $\psi_i$ and integration over the domain $\Omega$, resulting in

$$\int_\Omega \{(\delta_{im} - u_{i,m})S_{mj}\}\psi_{k,j}\,dV - \int_\Omega f_i\psi_k\,dV - \int_{\partial\Omega} g_i\psi_k\,dS = 0 \tag{8}$$

where also the boundary conditions from Eq. (2) are included.

Then, the discretization of the domain $\Omega$ is constructed using finite elements such as tetrahedra. Each element is equipped with a complete set of shape functions $\{\phi_i\}$ which are used for two purposes. First, the unknown function $u$ is interpolated as $u \cong \sum_i \phi_i \bar{u}_i$ where $\bar{u}_i$ are scalar coefficients representing the unknown variables. Further, the shape functions are used also as the weight functions $\psi_i$ in the weak form given by Eq. (8). Putting this together, the weak form can be reformulated over a typical element $\Omega^e$ resulting in a local system $\mathbf{K}^e(\bar{\mathbf{u}}^e) = \mathbf{f}^e$ where $\mathbf{K}^e$ is nonlinear local mapping between the actual displacement $\bar{\mathbf{u}}^e$ and response forces $\mathbf{f}^e$ for the element $e$. Global system $\mathbf{K}(\bar{\mathbf{u}}) = \mathbf{f}$ is then assembled from the contributions of the local elements.

The system is nonlinear due to the nonlinearity of the underlying PDE and cannot be solved directly. In this paper, iterative solution method combining the incremental loading and Newton–Raphson methods is considered. First, let's assume that the deformation for some given load $\mathbf{F}$ is to be computed. Then, an incremental loading path $\{(\mathbf{0} = \mathbf{F}_{(0)}, \mathbf{F}_{(1)}, \ldots, \mathbf{F}_{(M)} = \mathbf{F}\}$ is constructed and sequence of nonlinear systems $\mathbf{K}(\bar{\mathbf{u}}_{(i)}) = \mathbf{F}_{(i)}$ with initial estimation $\bar{\mathbf{u}}_{(i-1)}$ is solved for $i = 1, \ldots, M$. The solution in each step is obtained by Newton–Raphson method: having some estimation vector $\bar{\mathbf{u}}^{(j)}$ of the solution in the $j$-th iteration of the method, correction $\Delta^{(j+1)}$ is computed solving linear system and applied to obtain a new estimation of $\overline{\mathbf{u}}^{(j+1)}$

$$\mathbf{K}'(\overline{\mathbf{u}}^{(j)})\Delta^{(j+1)} = \mathbf{f} - \mathbf{K}(\overline{\mathbf{u}}^{(j)}) \tag{9}$$

$$\overline{\mathbf{u}}^{(j+1)} = \overline{\mathbf{u}}^{(j)} + \omega\Delta^{(j+1)} \tag{10}$$

where $\mathbf{K}'$ is global tangent stiffness matrix which can be again assembled from the local tangent stiffness matrices $\mathbf{K}'^{\mathbf{e}}$ computed for each element $\Omega^e$. The scalar factor $\omega \in (0,1]$ is obtained by line-search procedure in order to improve the convergence of the method by restricting the step size in each iteration so that the magnitude of decrease is maximized in each step of the Newton–Raphson method [19]. There are two criteria for the Newton–Raphson method to stop: first, if the condition $\|\mathbf{f} - \mathbf{K}(\overline{\mathbf{u}}^{(j)})\| < \varepsilon$ holds in $j$th iteration for some chosen $\varepsilon$, the resulting solution converges and is considered to be a valid data. Second, the iteration process is also stopped after $N$ maximum iterations also if the condition above does not hold. In this case, the calculation is nonconvergent and the data are considered to be invalid.

It is important to emphasize that before the solution to Eq. (9) takes place, the Dirichlet boundary conditions (BC) must be imposed. Basically, there are two main types: homogeneous and nonhomogeneous boundary conditions. In the first case, the zero displacements are prescribed for some components of the unknown vector $\overline{\mathbf{u}}$, i.e. $\overline{u}_i = 0$ for some $i$. These conditions fixing the body in the space are essential for the uniqueness of the solution. Besides, nonhomogeneous conditions $\overline{u}_j = b_j$ can be prescribed for some components of the vector $\overline{\mathbf{u}}$. The nonhomogeneous conditions can be regarded as another "input" of the method, since although no external forces are applied, the body gets deformed if nonzero deformation is prescribed for some part of the body (e.g. one node).

There are several techniques for imposing the Dirichlet boundary conditions such as elimination, penalization, and Lagrange multipliers. Among the three, the Lagrange multipliers can be used straightforwardly for obtaining the response of a node with prescribed displacement. Since the force is of a great importance for our application, the boundary conditions are imposed via Lagrange multipliers. The method based on the Lagrange multipliers is also utilized for the implementation of the incompressibility conditions introduced in the previous section. In this case, the system is augmented and the resulting multipliers can be understood as pressure applied to the nodes of the mesh.

## 3. Integration of soft tissue material properties into model

One of the main obstacles in developing realistic soft tissue models is the lack of data on the material properties of live organ tissues. Measuring and characterizing in vivo organ properties is a highly challenging task, but is a requirement for development of realistic surgical simulators. Soft tissue models with incorrect material properties will adversely affect training in VR-based surgical simulator systems. The research on tissue mechanics is extensive, but most of the earlier experiments took place in a laboratory environment (in vitro studies) under well-defined boundary and loading conditions. Typically, tissue samples taken from an organ of interest are transferred to a laboratory in a chemical solution for measurement. Because researchers carefully decide on the sample geometry and experimental conditions in advance, they can easily obtain stress and strain values from the measurement data. However, mechanical properties of soft tissues change with time and the results obtained through in vitro measurements do not represent actual tissue properties.

We developed a robotic indenter for minimally invasive measurement of live tissue properties in a living body [20]. This system includes a robotic arm (Phantom haptic device from Sensable Technologies, model 1.0), a force sensor (Nano 17 from ATI Industrial Automation), and a long probe that has a round tip with a 2 mm radius. Using the robotic indenter, we performed static indentation experiments on the liver of three pigs and successfully measured the nonlinear material properties of pig liver. An effective elastic modulus of pig liver was estimated from the static indentation data using the linear elastic contact theory and the small deformation assumption. In addition, an inverse FE solution was developed using ANSYS FE package to estimate the optimum values of nonlinear hyper-elastic material properties of pig liver via iterations. Hyper-elastic behavior of liver was modeled via 2-term Mooney–Rivlin strain-energy function defined by Eq. (7) in Section 2.1 and the material coefficients $C_{01}$ and $C_{10}$ are estimated from the inverse solution.

In this study, the linear elastic modulus of pig liver estimated from the static indentation data ($E = 15.48$ kPa) is utilized in the linear St. Venant material model to calculate the Lamé coefficients and the nonlinear material coefficients estimated via the inverse FE solution ($C_{01} = 1.28$ kPa, $C_{10} = 1.3$ kPa) are utilized in Mooney–Rivlin material model.

## 4. Visio-haptic simulation based on precomputations

### 4.1. Simulation parameters

In this section, the parameters affecting our visio-haptic simulations are introduced. In our simulations, we consider a displacement-driven interaction. The response forces displayed to a user of the system through a haptic device is calculated based on the position of the haptic interface point (HIP) in 3D space [21,22]. From the computational point of view, this implies that the position is taken as the input and force is calculated and returned as the output of the computations.

Let us assume a deformable body with a domain $\Omega$ that has been discretized resulting in the mesh $\mathcal{M}$ given by the set $\mathcal{V}$ of the vertices (nodes) and set $\mathcal{E}$ of the elements. First, the behavior observed during the interaction is determined by physical parameters. These are mainly the material coefficients defined for the selected model. Besides, there can be other parameters representing some external conditions; for example, the magnitude of an external force (e.g. gravitation) can be considered as physical parameter of the simulation. Second, there are geometric parameters which are related to the mesh of the deformable body. The first geometric parameter is represented by a set $\mathcal{F} \subset \mathcal{V}$ of surface nodes which are fixed in space during the interaction. From the numerical point of view, the set $\mathcal{F}$ defines the homogeneous Dirichlet conditions setting the components of the nodes in $\mathcal{F}$ to zero.

As the displacement driven interaction is being considered, there is at least one surface node, for which a nonzero displacement is prescribed. In this paper, single-point interaction is studied: when the haptic interaction point (HIP) penetrates into the object, a single active node $\mathcal{A}$ on the surface of the body is selected. During the simulation, the active node is associated with the components of its actual displacement. These values represent the last parameter which is, unlike the parameters introduced before, updated in each iteration of the haptic loop depending on the actual position of HIP. The prescribed displacement of $\mathcal{A}$ is therefore the control parameter of the simulation and from the numerical point of view, it can be regarded as nonhomogeneous Dirichlet condition.

Ideally, the computational model of the single-point static-equilibrium interaction works as follows: given the mesh of the deformable body, all the physical parameters of the model are specified together with the set $\mathcal{F}$ of the fixed vertices. Then, the interaction starts and in each iteration of the haptic loop, the instant position of HIP is acquired and the collision detection between HIP and the geometric surface model is performed. If the collision occurs, the surface node which is closest to the HIP position is selected as active and it is coupled with the HIP. From now on, in each iteration of the haptic loop, the control parameter (i.e. the actual position of the active node) is updated and the corresponding static equilibrium is calculated resulting in the overall deformation of the body and response force.

It is important to recall that the calculation of the static equilibrium of a complex body using FEM and prescribed displacement as the control parameter is a computationally expensive iterative process as shown in the mathematical background section. These iterations cannot be performed inside the haptic loop running at an update rate high enough for stable haptic interaction. Therefore, an approach based on precomputation of configuration spaces is presented for real-time interaction in the following section.

### 4.2. Configuration spaces: discretization and approximation

After specifying the parameters of the interaction, notion of configurations and configuration spaces are now introduced.

For a given set $\mathcal{F}$ of the fixed nodes and active node A, the actual state of the simulation can be completely described by configuration $C$ consisting of the current position $\mathbf{p}$ of HIP (the control parameter), vector $\mathbf{u}$ of displacements and force $\mathbf{h}$ acting on the active node.

As the control parameter changes from $\mathbf{p}$ to $\mathbf{p}'$, the new reaction force $\mathbf{h}'$ and deformation $\mathbf{u}'$ of the body are computed so that the static equilibrium is restored. In other words, a transition from the previous configuration $C$ to the new configuration $C'$ occurs: $C \rightarrow C'$. The set of all possible configurations, computed while having node $\mathcal{A}$ as active, is denoted as $\mathcal{S}_{\mathcal{A}}$. Now, the haptic interaction can be regarded as traveling through configuration space $\mathcal{S}_{\mathcal{A}}$ where each step is defined by a transition from one configuration to another.

Obviously, the configuration space $\mathcal{S}_{\mathcal{A}}$ is continuous and infinite. The main idea of the method is to discretize $\mathcal{S}_{\mathcal{A}}$ by constructing a finite subspace $\mathcal{D}_{\mathcal{A}} \subset \mathcal{S}_{\mathcal{A}}$ that can be efficiently precomputed and moreover, each configuration $C \in \mathcal{S}_{\mathcal{A}}$ can be approximated by some fast procedure using only the data from $\mathcal{D}_{\mathcal{A}}$. Thus, the approach proposed in this paper can be summarized as follows:

*Off-line precomputation phase*: For a given FE mesh, all the physical parameters are specified and both the set $\mathcal{F}$ and the active node $\mathcal{A}$ are selected. Then, a set of points $\mathcal{G}_{\mathcal{A}}$ is chosen around the rest position of the active node $\mathcal{A}$ and for each such a point $\mathbf{g} \in \mathcal{G}_{\mathcal{A}}$, the active node is displaced to the position $\mathbf{g}$ and the corresponding configuration is computed and stored. Details on this process are presented in Section 4.3.

*On-line interaction phase*: The precomputed configurations from the discrete set $\mathcal{D}_{\mathcal{A}}$ are used to approximate an arbitrary configuration $\tilde{C}$ associated with the current position $\mathbf{p}$ of HIP. The approximation is calculated by interpolation of the precomputed data which is fast enough to be performed inside the haptic loop. The details of the interpolation phase are presented in Section 4.4.

### 4.3. Off-line precomputations

In this section, we discuss the methods for the selection of points in $\mathcal{G}_{\mathcal{A}}$ and the construction of $\mathcal{D}_{\mathcal{A}}$. The algorithm presented for the construction of $\mathcal{D}_{\mathcal{A}}$ is based on the assumption that the solution of the boundary problem solved in each step of the simulation is unique. This assumption is based on the ellipticity of the boundary problem being solved in each step of the simulation. This assumption combined with the single-point interaction implies that there is at most one configuration $C$ corresponding to the particular position $\mathbf{p}$ of the HIP.

By virtue of the assumption about the uniqueness of the solution, it is sufficient to define the discretization structure $\mathcal{G}_{\mathcal{A}}$ as a *uniform grid* of points surrounding the rest position $\mathbf{x}_0^{\mathcal{A}}$ of the active node $\mathcal{A}$ which coincides with one of the grid points. So for $i, j, k \in \mathbb{Z}$, the grid is defined as

$$\mathcal{G}_{\mathcal{A}} = \{\mathbf{g}_{ijk} \in \mathbb{R}^3 \,|\, \mathbf{g}_{ijk} \le r_G \wedge \mathbf{g}_{000} = \mathbf{x}_0^{\mathcal{A}}\} \qquad (11)$$

There are two parameters of the grid. The first one is represented by the radius $r_G$ determining how far the active node is going to be displaced from the rest position. The second parameter sets the density of the grid, i.e. the distance between two adjacent points of the grid.

The construction of the corresponding set $\mathcal{D}_{\mathcal{A}}$ can be defined recursively as follows.

1. For the point $\mathbf{g}_{000}$ corresponding to the rest position of the active node, store the zero configuration $C_{000}$ where all the displacements of the nodes as well as the response forces are set to zero.
2. If a configuration $C_{ijk}$ has been already computed and stored in the point $\mathbf{g}_{ijk}$, compute the transitions $C_{ijk} \rightarrow C_{i'j'k'}$ to all adjacent positions $\mathbf{g}_{i'j'k'}$, such that $\|i - i', j - j', k - k'\| \in \{1, \sqrt{2}, \sqrt{3}\}$ provided $C_{i'j'k'}$ has not been already computed. This step can be regarded as a *configuration expansion*.
   The particular transition $C_{ijk} \rightarrow C_{i'j'k'}$ is constructed by the incremental loading technique: the path between $\mathbf{g}_{ijk}$ and $\mathbf{g}_{i'j'k'}$ is split into $M$ steps and $C_{ijk}^0$ is set to $C_{ijk}$. Then, the path is traversed by executing Newton–Raphson method to compute the configuration $C_{ijk}^m$ from the initial state $C_{ijk}^{m-1}$ for $m = \{1, \ldots, M\}$.
   The intermediate configurations can be stored as well since they can be utilized by the interpolation methods implemented in this study to calculate reaction forces.
3. Repeat the step 2 until a corresponding configuration $C_{ijk}$ is computed and stored for each point $\mathbf{g}_{ijk} \in \mathcal{G}_{\mathcal{A}}$.

It must be emphasized that the configuration space will not cover all the points from the grid $\mathcal{G}_{\mathcal{A}}$ (i.e. for some of the points in the grid, the corresponding configurations will not be computed). There are two main reasons for this:

1. Haptic devices have a physical limitation concerning the maximal force $\mathbf{f}_{max}$ that it can display. This implies that if the reaction force $\mathbf{f}$ in some configuration $C_{ijk}$ exceeds $\mathbf{f}_{max}$, the configuration is not reachable since the force limitation of the device does not allow the user to reach this position by HIP. Therefore, such a configuration is not expanded any more and the configurations corresponding to the points beyond $\mathbf{g}_{ijk}$ are not computed.
2. The construction of each step within a transition $C_{ijk} \rightarrow C_{i'j'k'}$ is an iterative process which can fail numerically. This can happen for several reasons; either the deformation is too large to be modeled physically, or some elements of the mesh get degenerated causing the distortion of the mesh or the chosen material is not suitable for that type of deformation. If such a *failure point* occurs on the path from $\mathbf{g}_{ijk} \rightarrow \mathbf{g}_{i'j'k'}$, the configuration $C_{i'j'k'}$ is not inserted to the state space (i.e. there is no configuration for the point $\mathbf{g}_{i'j'k'}$).

### 4.4. On-line interaction

In this section, the interpolation of the reaction force $\mathbf{h}$ which is computed during the real-time haptic interaction phase is explained. The nodal displacements $\mathbf{u}$ are computed in the same manner for the visualization purposes at a lower refresh rate. Each component of $\mathbf{h}$ is interpolated separately using the precomputed force data. In the following, two different methods are considered: polynomial and radial-based interpolation (see survey [23]). For each of them, linear and cubic versions are employed.

The polynomial interpolation works with uniformly distributed grid data, i.e. only the precomputed configurations $C_{ijk}$ stored in the points of the grid $\mathcal{G}_{\mathcal{A}}$ can be utilized. For the simplest case of the linear interpolation, the cell of the grid $\mathcal{G}_{\mathcal{A}}$ where the currently active node $\mathcal{A}$ is located can be identified based on the position of $\mathcal{A}$ ($\mathbf{p}_{\mathcal{A}}$). Afterwards, the components of the force vector can be easily calculated by tri-linear interpolation using the eight corners of the cell. The technique is illustrated in Fig. 1(a). Clearly, the tri-linear interpolation is fast, however, it becomes unreliable as the nonlinearities in the model dominate the force response. One can use a finer grid to solve this problem, but this approach increases the number of precomputations significantly. It is also important to emphasize that the interpolation within a cell can be computed provided all eight nodes of the cell are available. Therefore, as soon as the current position $\mathbf{p}_{\mathcal{A}}$ gets out of the grid, the configuration cannot be approximated well.

The next alternative within the frame of the polynomial method is *tri-cubic interpolation*. In this case, besides the cell containing the actual position $\mathbf{p}_{\mathcal{A}}$, eight additional neighboring cells, resulting in a *super-cell* with 64 grid points, are used for the interpolation (see Fig. 1(b)). In this case, computations are more demanding, but still can be done within the haptic loop. In the case of tri-linear interpolation, the tri-cubic interpolation is reliable only if all the configurations for the 64 points of the super-cell is already available. For a given grid of precomputed configurations, the tri-cubic interpolation may not work close to the border of the grid or for the case when the grid points are missing due to the nonconvergent calculations. Therefore, it cannot be applied to the regions close to the boundary of the grid where sufficient data are not available.

Regarding the complexity of the polynomial interpolations, the number of floating-point operations for both tri-linear and tri-cubic polynomial interpolations are independent of the size of the grid: in the former case, about 250 floating-point operations are needed to compute one component of the force, whereas 2000 operations are necessary for the tri-cubic method. When comparing to the performance of today's processors, the interpolation of the force vector can be easily achieved within the haptic loop.

For further improvement in the accuracy of the approximation, a *radial-basis function* (RBF) is employed for the interpolation (see Fig. 1). It is suitable for irregularly scattered data, so beside the configurations $C_{ijk}$ stored in the points of the grid $\mathcal{G}_{\mathcal{A}}$, the intermediate states $C_{ijk}^m$ can be utilized as well. For example, having a set of positions $\mathbf{x}_i$ in space together with the corresponding associated values $y_i$ stored in those positions, the interpolated value $y$ at some position $\mathbf{x}$ is computed as

$$y = \sum_{i=1}^{N-1} w_i \phi(\|\mathbf{x} - \mathbf{x}_i\|) \tag{12}$$

where $\phi$ is a user-defined function known as the *kernel*. In our study, $\phi(r) = |r|$ and $\phi(r) = r^3$ are used for linear and cubic interpolation, respectively. The interpolation procedure given above is used for each component of the force and displacement vectors. The weights $w_i$ can be calculated during the off-line phase of the computation by solving the linear system which is assembled after substituting the components of the force vector $\mathbf{h}$ into Eq. (12). Similarly, the nodal displacements can also be interpolated using Eq. (12).

Therefore, the proposed approach consists of two consecutive phases—computation of the weights and interpolation. Each phase of the computation has different complexities. Let us first focus on the interpolation phase. Comparing to the tri-linear and tri-cubic interpolation, the computational complexity of this phase linearly depends on the number of nodes in the grid. In other words, it is $\mathcal{O}(\sqrt[3]{N})$ for the three-dimensional grid $r_G$.

The interpolation phase can be easily executed in real time for grids having number of grid points varying from hundreds to thousands depending on the performance of the computer running the haptic loop. On the other hand, the computational complexity of the first phase (computation of weights) is $\mathcal{O}(N)$ and it is computationally much more demanding. In the actual setting, the radius of influence of each grid point is infinite for the force interpolation. Therefore, a dense system must be solved in order to compute the RBF weights. Nevertheless, this can be done efficiently using LU decomposition during the precomputation phase, adding only a small amount of computation time to the total time spent for the precomputations. The radial basis function can be used for the interpolation of points close to the border of the uniform grid. In fact, it can be used even when there are some configurations missing. However, the accuracy of the solution is not guaranteed anymore and the results of the approximation may become physically invalid.

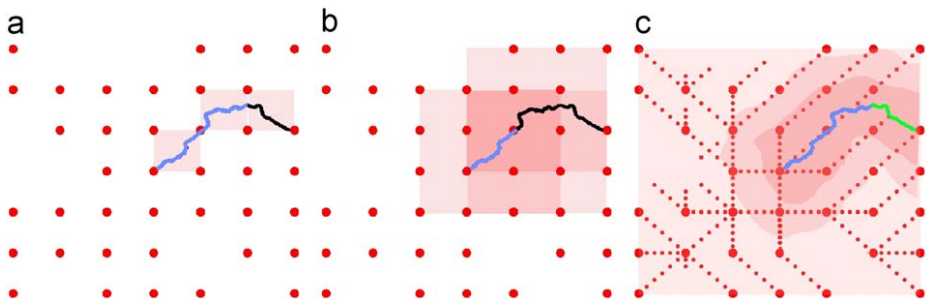

**Fig. 1.** A 2D illustration of the three interpolation methods discussed in this paper: (a) bi-linear interpolation (b) bi-cubic interpolation and (c) interpolation using radial-basis function.

# 5. Implementation and computations

## 5.1. Distributed implementation of the space precomputation

First, we briefly focus on the implementation of the precomputation phase. As stated before, all the deformations considered in this paper have been computed using finite element method. For this purpose, a freely available finite element library GetFEM [24] has been employed. The solution of the nonlinear FE equations for small changes of the control parameter (i.e. the prescribed displacement of a node) is obtained by incremental loading. In each step, the Newton–Raphson method with line search is applied to minimize the residual. In each iteration of the Newton method, the linearized system is solved using the MUMPS linear solver [25].

The construction of the configuration space has been implemented recursively as described in the Section 4.3. In order to speed up the precomputations, the independent calculations are executed in a distributed manner using a cluster of computers. A centralized approach is used for the implementation; a central client process working as a very simple scheduler distributes the work among the servers. There are two tables maintained by the scheduler. First, it is the table of idle servers which are available. Initially, all the servers are inserted to the table. Second, it is the table of all the configurations which are to be computed. Each is marked by exactly one of the following tags:

- *unknown*: the configuration has not been computed so far, only its location within the grid is known;
- *expandable*: the configuration has been successfully computed and can be used as initial estimation for further computations;
- *terminating*: the configuration has been successfully computed, but it is not reachable by the haptic device, so it is not necessary to perform further computations;
- *failed*: the computation of the configuration has failed due to the convergence issues.

Initially, the zero configuration corresponding to the rest position of the active node is marked as expandable and all other configurations are marked as unknown. The configuration space is then constructed as follows: if there is at least one idle server, the scheduler chooses some unknown configuration $C'$ for which there is an expandable configuration $C$ such that $C$ and $C'$ are adjacent in the grid. Afterwards, the computation of the transition $C \rightarrow C'$ is assigned to the idle server which is removed from the table of idle servers.

As a result of the transition, a three different tags can be assigned to the configuration $C'$. First, if the iterative transition process does not fail, then the configuration $C'$ is valid and the magnitude of the estimated reaction force is compared to the force limit $f_{max}$. If it is less than the limiting value, the configuration $C'$ is marked as expandable since it is reachable during the interaction and other configurations can be computed using $C'$ as an initial estimation. In the opposite case, the configuration is marked as terminating, because it is not necessary to expand it further since it is not reachable by the haptic device. In both cases, the configuration $C'$ is stored to be used later for the interpolation process. Second, if the transition process fails, the configuration $C'$ is marked as failed. In this case it is not stored, since it cannot be used by the interpolation process. For all the cases enumerated above, the scheduler is notified by the server about the results of the transition computations, so that the server can update the tables: the tag of the configuration $C'$ is updated according to the results of the computations and the server is set to idle. This process is repeated until all the configurations in the grid are marked as either expandable, terminating or failed.

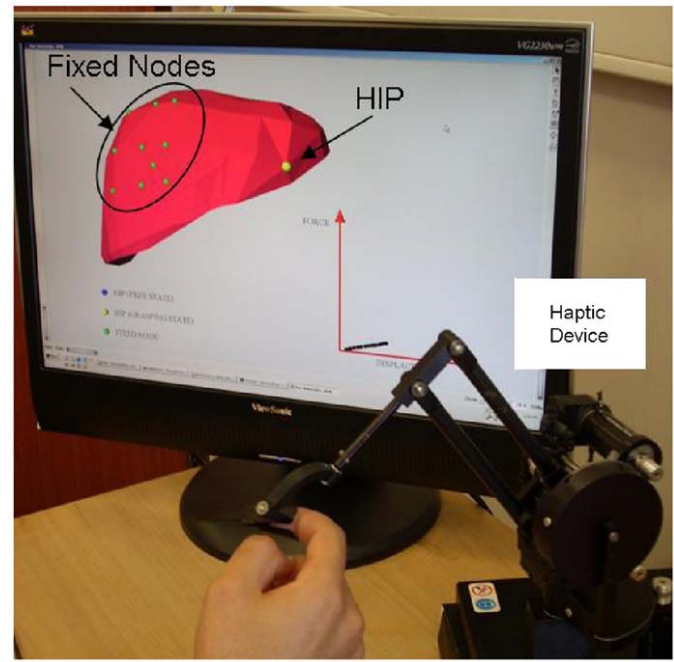

**Fig. 2.** The hardware components of our simulation system.

## 5.2. Implementation of the space approximation

After giving the details of the precomputation phase, we now focus on the implementation of the interaction phase. We have simulated the real-time visio-haptic interactions between a point (i.e. HIP) and the 3D surface model of human liver (made of triangular patches). The hardware components of our simulation system include a computer monitor for visual display of deformations and a haptic device for feeding the interaction forces back to the user (Fig. 2). The underlying code is written in MS Visual C++ environment, the graphical rendering of the liver model and the visual deformations are displayed using Open Inventor (a scene graph API) [26,27], and the haptic feedback to the user is provided via PHANToM haptic device (SensAble Technologies) using GHOST v.4.0 driver.

Our real-time computational architecture for simulating nonlinear tissue response consists of two threads running asynchronously: haptic and visual threads. The haptic thread, updated at 1 kHz, acquires the new position of the haptic probe as the user manipulates the probe, performs collision detection, determines the active node, and calculates the collision response based on the interpolation of the precomputed data stored at the grid points surrounding the active node. In the current setting, the HIP instantly snaps to the nearest vertex. The visual thread, updated at 30 Hz, graphically renders the haptic interaction point and the deformations of the 3D liver model.

## 5.3. Model and computations overview

All the experiments discussed in the next paragraph of this section have been performed on the 3D model of human liver obtained from the INRIA repositories. The model has been meshed by TetGEN mesh generation tool [28], resulting in two meshes with 1777 elements (501 nodes) and 10,280 elements (2011 nodes), respectively. For the experiments, two different sets of fixed points $\mathcal{F}_1$ and $\mathcal{F}_2$ are tested, fixing the area on the left and top parts of the organ, respectively. Further, 10 different active nodes are selected for the testing purposes (see Fig. 3(a)).

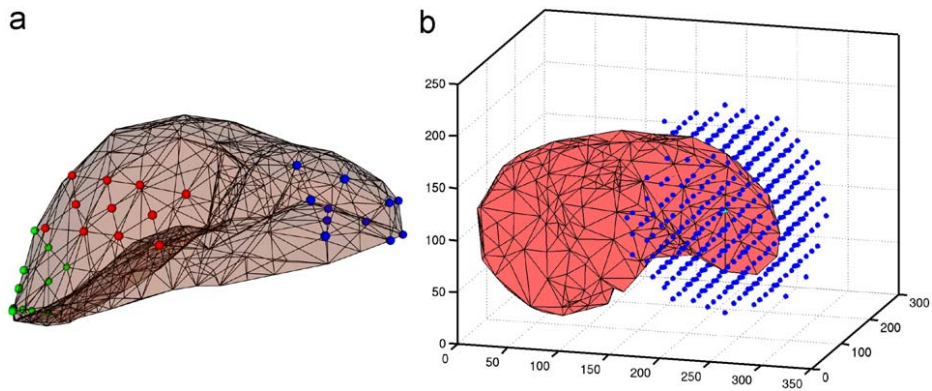

**Fig. 3.** (a) The liver model used for the experiments. The nodes are colored according to the functionality: the green and red nodes belong to the fixed-node sets $\mathcal{F}_1$ and $\mathcal{F}_2$, respectively. The blue nodes belong to the set of the active nodes. (b) Illustration of a grid used for the precomputation of configuration spaces. (For interpretation of the references to color in this figure legend, the reader is referred to the web version of this article.)

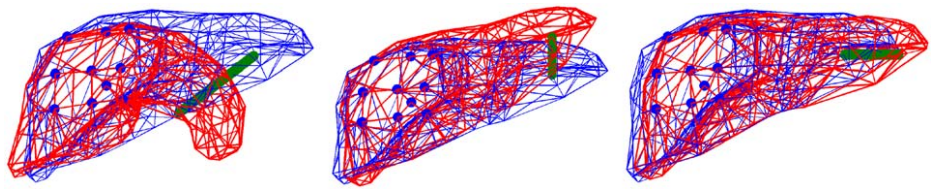

**Fig. 4.** Three frames from the real-time FE simulation showing various types of deformations. The rest pose of the liver is depicted by blue, the deformation caused by displacement path of a node (green) is depicted by red. (For interpretation of the references to color in this figure legend, the reader is referred to the web version of this article.)

Regular and random loading paths are generated and the corresponding configuration spaces are computed. Regarding the regular paths, 26 paths are tested for 10 active nodes with two sets of fixed nodes ($\mathcal{F}_1$ and $\mathcal{F}_2$) so in total of 520 paths ($26 \times 10 \times 2$) are tested for each material type. Each path starting in some active node $\mathcal{A}_i$ is 10 cm long and the length of each step in the incremental loading is 1 mm to improve the convergence of the Newton–Raphson method. The construction of a single path takes from 15 to 30 min depending on the type of material law and convergence rate. Since more than 150,000 configurations have been constructed in total, the generation of regular paths has been performed on a cluster having 32 CPU cores (Intel Xeon running at 3.0 GHz). It takes about 60 h to compute all configurations. The results are stored on shared disk arrays for inspection and interpolation later. Regarding the random paths, about 5000 paths are tested for each material type. Each path consists of 200 steps and the incremental step size is 1 mm. In total, about 3 million configurations are constructed. The computation of random paths takes about 120 h on cluster of 48 CPUs (Intel Xeon running at 3.0 GHz). Four uniform grids with different densities are generated and the corresponding configurations are computed. Each grid is enclosed in a sphere with a radius of 10 cm for each active node being the center of the sphere. The points inside the grid are distributed regularly dividing the volume of the sphere into cubic cells. The cell sizes are 20 mm, 14.3 mm, 10 mm and 6.667 mm, the corresponding grids have 514, 1418, 4168 and 14,146 points respectively. The grid with 514 points is illustrated in Fig. 3(b) and three frames of the simulation are shown in Fig. 4.

## 6. Evaluation and discussion

### 6.1. Behavior of the model

In the following, the behavior of the liver model is briefly studied using force–response plots obtained from the application of the regular loading paths introduced in Section 5.3. First, the stability of the computations is analyzed, as the nonlinear systems are considered. Then, the influence of active node and boundary conditions on the force displacement is investigated.

The detection of the nonconvergent computations during the construction of the configurations is crucial for the evaluation of the method presented in this paper, since the interpolation cannot be applied successfully to invalid data produced by nonconvergent computations (see Section 2.2). The tests using the regular paths show that the convergence problems are mostly encountered for the St. Venant–Kirchhoff material. In this case, 53% of the paths are nonconvergent compared to less than 5% in Mooney–Rivlin material. The visual comparison of linear elasticity, St. Venant and Mooney–Rivlin materials with nonlinear strain tensor is given in Fig. 5.

Further investigation reveals that mainly the paths causing compression-like deformation are problematic. The history of the mesh deformation is then studied more precisely for the nonconvergent paths. It turns out that despite the fact that incompressibility is imposed implicitly for the St. Venant material by setting Poisson ratio to 0.49, some elements are degenerated during the loading process. In order to solve this problem, the explicit incompressibility conditions (see Section 2.1) are imposed also for the St. Venant material resulting in much more stable behavior with 95% of convergent paths. The St. Venant material equipped with the incompressibility conditions is called as modified St. Venant material in the paper.

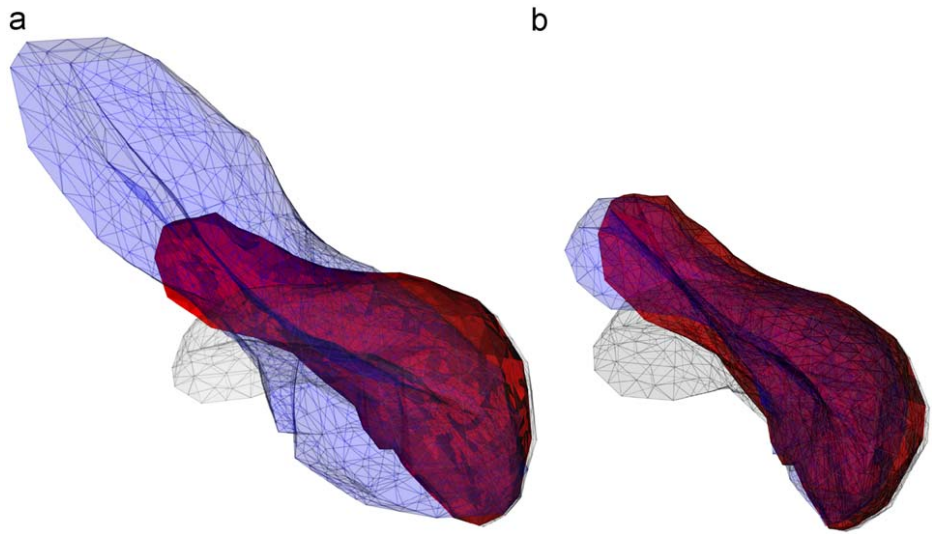

**Fig. 5.** Visual differences for various models. On both figures, gray and red colors are used for the undeformed mesh and the deformed mesh modeled using the Mooney–Rivlin material. The deformation shown in blue is obtained by using (a) linear elastic equations (notice the excessive volume change) and (b) St. Venant–Kirchhoff material with nonlinear strain tensor (only a small difference is observed w.r.t. the Mooney–Rivlin material). (For interpretation of the references to color in this figure legend, the reader is referred to the web version of this article.)

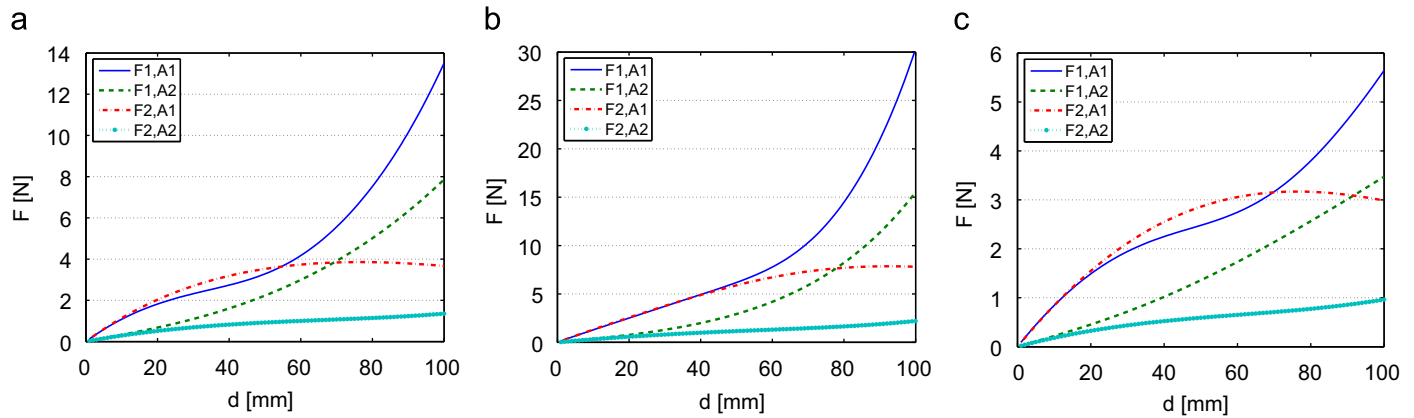

**Fig. 6.** Force–displacement curves for the combinations of the fixed nodes ($\mathcal{F}_1$, $\mathcal{F}_2$) and active nodes ($A_1$, $A_2$). The results are depicted for three materials: (a) St. Venant–Kirchhoff, (b) modified St. Venant–Kirchhoff and (c) Mooney–Rivlin.

Location of the fixed nodes and active node are used as the control parameters while simulating the behavior of liver model for different material laws. For the evaluation, two sets of fixed nodes $\mathcal{F}_1$ and $\mathcal{F}_2$ and two choices of active nodes $A_1$ and $A_2$ are used (the former located closer to the gravity center of the body, the latter closer to the right "tip" of the liver). The force–displacement responses of the three materials for different combinations of the fixed nodes and the active nodes are given in Fig. 6. When comparing the shape of the curves within each graph, it is clear that selection of the boundary conditions (i.e. fixed and active nodes) plays important role in force response. First, the concave curves have been observed for the case when the set of nodes $\mathcal{F}_2$ on the top of the liver was fixed. This selection of the fixed nodes is quite restrictive resulting in local deformations. Hence only the elements around the active node are deformed to follow the prescribed displacement. On the other hand, for the set $\mathcal{F}_1$ the liver can be deformed globally, since both active nodes are located on the opposite side of the liver w.r.t. the set $\mathcal{F}_1$. This results in convex force-response. It is observed that the magnitude of the response force seems to be determined by the location of the active node w.r.t. the liver geometry. For example, the displacement of the node $A_1$ results in larger response force than that of $A_2$ (for the same boundary condition $\mathcal{F}_1$), since it is closer to the gravity center. Putting it all together, the force–displacement curves recorded during the large number of the experiments with regular loading paths show the importance of the boundary conditions, i.e. location of the fixed and active nodes, on the force response of the soft body.

**Table 1**
Wall time of single Newton iteration showing the length of the assembly phase (computation of the stiffness and tangent stiffness matrices), LU-based solution phase and the total (sum of assembly and solution) in seconds.

| No. | 1 | 2 | 3 | 4 | 5 |
|---|---|---|---|---|---|
| Material | SV | MR | SV | MR | MR |
| Order | Linear | Linear | Linear | Quadratic | Linear |
| #nodes | 501 | 501 | 2011 | 501 | 2011 |
| #elements | 1777 | 1777 | 10,270 | 1777 | 10,270 |
| #equations | 1539 | 2040 | 6069 | 12,323 | 8080 |
| Assembly (s) | 0.49 | 1.05 | 3.22 | 3.38 | 8.15 |
| Solve (s) | 0.05 | 0.23 | 0.71 | 3.18 | 1.74 |
| Total (s) | 0.54 | 1.28 | 3.93 | 6.56 | 9.89 |

## 6.2. Construction of configuration spaces

Before discussing the computational cost of constructing configuration space, we first present the computational time required for one Newton–Raphson iteration employing two different materials, St. Venant–Kirchhoff (SV) and Mooney–Rivlin (MR) and two meshes of different sizes: one with linear and another with both linear and quadratic elements (see Table 1). The computational times for the modified St. Venant material are close to those obtained for the Mooney–Rivlin material.

The computations have been executed sequentially on AMD Opteron Processor 250 (2 GHz) with 8 GB of physical memory. The execution times reported in Table 1 are the average of large number of iterations. The larger number of equations for the Mooney–Rivlin material is caused by the enforcement of the incompressibility conditions via the Lagrange multipliers. This also leads to longer time needed for the assembly process. In the case when line search technique is applied, additional time is required in each iteration. Since the wall time spent by a single Newton iteration is shown, it must be multiplied by the number of iterations needed for the computation of one configuration. Therefore, the total time of the Newton method including the line search can result in more than 1 min.

For this reason the configuration space is constructed using a distributed computing approach. All the experiments have been executed on 16 nodes of a cluster, each having 2× Dual Core AMD Opteron Processor 270 with 8 GB of physical memory running at 2 GHz, so in total, 64 processor cores (16 × 4) have been used for the precomputation and the data is stored on local disks for further processing. The lengths of the computations are shown in Fig. 7. The configuration space can be constructed quickly for the grid with 514 points, ranging from 2 to 28 min depending on the complexity of the model. The computational cost of constructing the grid with 1418 points is still acceptable (about 1 h for FE model having more than 10,000 elements). The finer grid (4168 points) is affordable mainly for simpler models (a linear FE model having 501 nodes is computed in less than 45 min), but more complex models may take about 3 h. Finally, the grid with 14,146 points is tested only with two simple FE models, confirming the cubic increase of the computational time due to the increase in number of grid points.

An important aspect of the presented work is the size of the data file, storing the precomputed configurations. For the coarser FE mesh with 1777 elements, the size of data file is 3 MB for the grid with 514 points and 81 MB for the grid with 14 146 grid points. Further, for the FE mesh with 10,270 elements, the corresponding sizes are 11 and 325 MB, respectively. Therefore, a data file storing the deformation configurations of a 3D object under the influence of a point load (i.e. a single active node) can be easily loaded into the memory of today's computers to be used for real-time computations later.

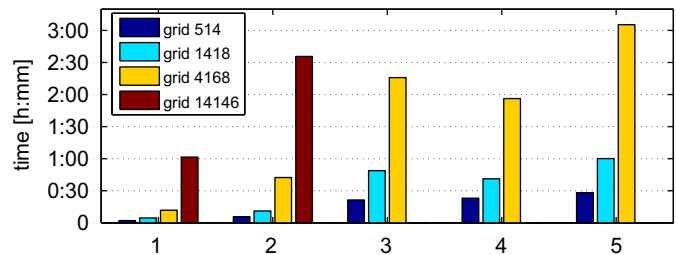

**Fig. 7.** The clock time needed to compute a single configuration space for the Mooney–Rivlin material using the FE mesh with 1777 elements. The computations were run on 64 CPU cores. Each group corresponds to one combination of parameters as indexed in Table 1.

## 6.3. Evaluation of the space reconstruction

In this section, the reconstruction process of the configuration space is validated. This is performed by comparing the configurations computed precisely with their counterparts obtained by the interpolation of the precomputed data. For this purpose, both the regular and the random loading paths are utilized: for each configuration $C$ computed during a particular loading, its counterpart $\tilde{C}$ is reconstructed by interpolation using the precomputed spaces. This reconstruction process is tested with four different interpolation methods: tri-linear and tri-cubic and radial-basis function with linear and cubic kernels. Moreover, each interpolation method is tested with four precomputed uniform grids having different number of points (514, 1418, 4168 and 14,146).

For the purpose of evaluating the interpolation accuracy, two error measures are defined and calculated for each interpolation method and the grid density. Both error measures are calculated using the components of the force ($x$, $y$ and $z$ in this case) instead of its magnitude in order to determine the accuracy more strictly. First, the absolute mean error $\overline{E}$ showing the average difference between the interpolated force component $\tilde{f}_{\tilde{C}}^{i}$ and precisely computed force component $f_{\tilde{C}}^{i}$ of each configuration along a path $\mathcal{P}$ is calculated as

$$\overline{E}_{\mathcal{P}} = \frac{1}{3|\mathcal{P}|} \sum_{i \in \{x,y,z\}} \sum_{C \in \mathcal{P}} |f_{\tilde{C}}^{i} - \tilde{f}_{\tilde{C}}^{i}| \qquad (13)$$

Similarly, the relative mean error $\overline{e}$ is computed as ratio of absolute mean error to the precisely computed force component $f_{\tilde{C}}^{i}$:

$$\overline{e}_{\mathcal{P}} = \frac{1}{3|\mathcal{P}|} \sum_{i \in \{x,y,z\}} \sum_{C \in \mathcal{P}} \frac{|f_{\tilde{C}}^{i} - \tilde{f}_{\tilde{C}}^{i}|}{|f_{\tilde{C}}^{i}|} \qquad (14)$$

When small forces (e.g. below 0.5 N) are interpolated, the relative error can increase rapidly, although the absolute difference is even bellow the resolution of the device or human haptic perception. For the case of larger forces (above 20 N), the absolute difference can be significant (over 1 N), but it is negligible w.r.t. the overall magnitude of the force.

To be able to present the details in a compact way, the absolute and relative errors introduced above are further averaged over all the computed paths resulting in mean absolute error $\overline{E}$ and mean relative error $\overline{e}$. The results for different interpolation methods are presented in Fig. 8.

The graph shows that the influence of the interpolation on the accuracy is significant. For example, the accuracy obtained by the RBF interpolation with cubic kernel using the coarsest grid with 514 points is comparable to the accuracy of the tri-linear interpolation using the finest grid with 14 146 points. The tri-cubic interpolation seems to show the best accuracy w.r.t. the absolute error. Nevertheless, it is important to emphasize that the tri-cubic interpolation can be used only if a supercell, composed of 64 grid points, is available. For example, for the grid having 514 points, only 31% of the configurations generated during the experiments can be interpolated using the tri-cubic method due to the missing data. Therefore, the utilization of tri-cubic interpolation is limited to a smaller area closer to the rest position of the active node. This reduces the absolute error since larger forces corresponding to the boundary areas are not included in this type of interpolation.

Besides presenting the average error for the different interpolation methods, the maximum error representing the worst case scenario is investigated using the maximum relative and absolute errors for each path denoted $e_{\mathcal{P}}$ and $E_{\mathcal{P}}$, respectively. The definition of both quantities is similar to Eqs. (13) and (14) except the summation signs are replaced with maximum signs.

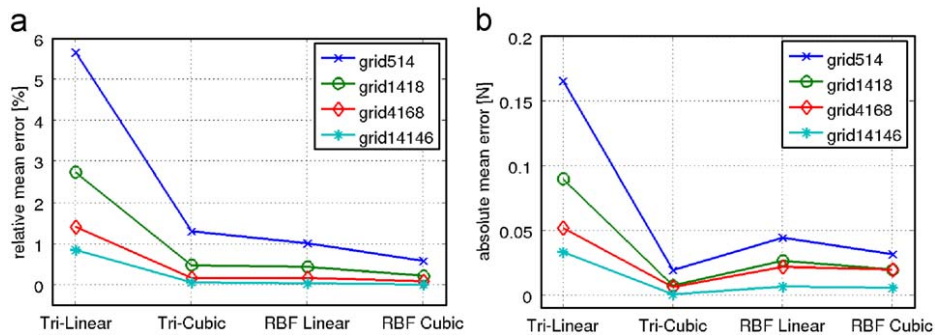

**Fig. 8.** Relative mean error $\bar{e}$ depicted in (a) and absolute average error $\bar{E}$ depicted in (b) for the various grids and interpolation techniques.

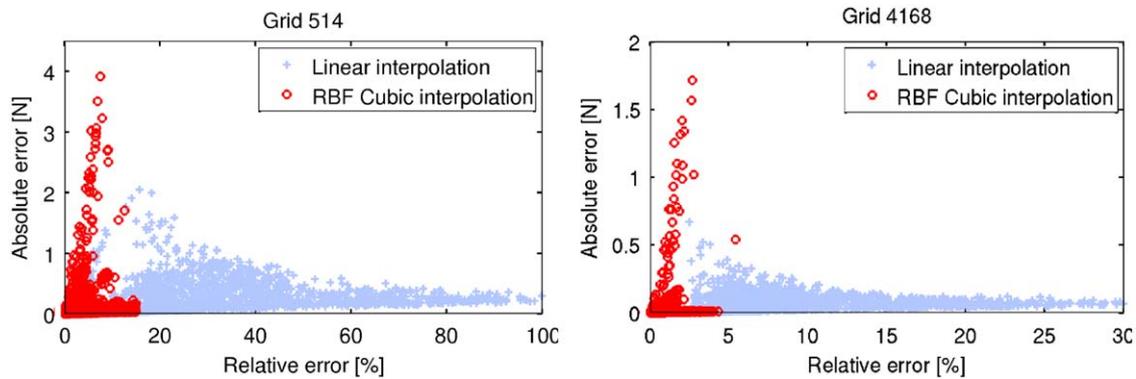

**Fig. 9.** Maximum errors of tri-linear and RBF-cubic interpolation: maximum error for each generated path is depicted by one circle and cross, respectively.

The results are shown in Fig. 9 where tri-linear and RBF cubic interpolations are presented for two grids having different densities. The graph shows that the maximum relative error (x axis) can be significantly reduced if RBF cubic interpolation is used instead of tri-linear interpolation. The figure shows that tri-linear interpolation on the dense grid (blue dots on the right-hand side) results in maximum relative error achieving 30%, whereas the RBF interpolation on the coarse grid (red dots on the left-hand side) results in maximum relative error under 20%. Moreover, the distribution of dots in both graphs show the relation between the absolute and relative errors. As the dots depicting the maximum errors are close to the axis, it can be concluded that although the relative error is large, the corresponding absolute difference is low or vice versa, the large absolute error occurs mainly when the absolute magnitude of the force is large and so the relative error is in fact low.

## 7. Discussion and future work

In the first part of the evaluation section, the influence of the simulation parameters such as material type and location of fixed and displaced nodes on the force and displacement response of a deformable, but static FE model was studied. It was concluded that the parameters significantly affect the behavior of the simulation if geometric and material nonlinearities are considered in the model. From the numerical point of view, the convergence of the computations is determined by the FE model, whereas from the physical point of view, the response of the material depends on the actual choice of the material law and boundary conditions. Therefore, the technique proposed in this paper does not depend on the complexity of the mathematical model and its parameters, as any computationally expensive calculations are performed off-line, independently from the real-time haptic interaction. For example, if the actual setting of the parameters results in slow convergence of the computations, this will not affect the stability of the haptic interaction. Moreover, if some configurations are not valid due to nonconvergence, they are excluded from the data set before the interaction phase, so they do not introduce additional error into the interpolation. Further, the configuration spaces for various models and parameters can be constructed during the precomputation phase, so the differences in the behavior of the tissue can be tested and compared on-line during the haptic interaction. Putting it all together, our approach allows the user to study various static deformation models and parameter settings without any influence on the haptic loop.

It is important to emphasize that although mainly the force response was studied in the paper, the approach covers both visual and haptic rendering: during the precomputation phase, the displacement vectors are generated together with the forces, so they can be used to approximate the actual deformation of the body. During the interaction, the displacement components for each visible node must be interpolated, nevertheless this can be safely done in the visual loop running at a refresh rate of 30 Hz. Moreover, since the visual accuracy is not as crucial as the haptic accuracy, the tri-linear interpolation of the nodal displacements is by far sufficient.

On the other hand, the main limitation of the proposed approach is the computational time required for the precomputations. Also, necessity of constructing the configuration spaces for each model and parameter setting can be inconvenient. Nevertheless, the length of the precomputation phase can be significantly shortened using a distributed computing approach suitable for computational grid architectures. As the algorithms does not depend on the network latency, it scales well also for large number of servers involved in the computations.

Concerning the reconstruction phase, the simple tri-linear interpolation can be acceptable, if implemented on a denser grid. The tri-cubic interpolation displayed better results even on the sparser grids, but this method can be used only when all the 64 points are available for the interpolation of a configuration. Nonetheless, these two interpolation methods can be used together so that the tri-cubic interpolation is used in the region close to the rest position of the active node, where sufficient data is available, whereas the tri-linear method is applied to the regions close to the border of the grid.

The RBF method showed to be the best choice when the accuracy is important, as it still can be computed within the haptic loop, provided that the weights are computed within the precomputation phase. It was shown even that the cubic variant of RBF applied to the sparsest grid gives better results in terms of the accuracy than the tri-linear method applied to the densest.

In this paper, a single point interaction is studied employing 3DOF haptic device. The extension of the approach towards multiple point interaction would result in significant increase in the number of precomputed configurations, since $k$ simultaneously controlled nodes would introduce a multiplicative factor of $\mathcal{O}(2^k)$. Nevertheless, the configuration space can be indexed by the position and orientation of the 3D tool interacting with the tissue to reduce the size of the configuration space. Then, the configurations for a given set of tool positions and orientations can be computed and stored during the precomputation phase and reused for the interpolation during the interaction. A preliminary study implementing this idea using a spherical tool made of multiple points can be found in [29]. This topic requires further investigation in the future.

Concerning the future work, we are currently working on a solution which does not depend on the time consuming precomputation phase. The idea is based on the fact that the motion of HIP controlled by the user is relatively slow. Therefore, if the trajectory followed by HIP is roughly estimated based on the previous positions of HIP, the possible configurations surrounding the trajectory can be computed in advance and the reaction force corresponding to the actual position of HIP can be interpolated from the precomputed subspace. Hence, the whole configuration space is not constructed again, but only a part of it is updated. Then, GPU accelerators can be used perhaps to generate the configurations in real time. We believe that this approach based on local configuration spaces can handle topological changes and dynamical effects, which are both of great interest in the area of surgical simulations.

## Acknowledgments

The first and last authors acknowledge the financial support provided by Ministry of Education, Youth and Sport of the Czech Republic under the research intent number 102/05/H050. The second and third authors acknowledge the financial support provided by TUBITAK under contract number MAG-104M283 and the student fellowship program BIDEP-2210. The access to the METACentrum computing facilities provided under the research intent MSM6383917201 is acknowledged.

## Appendix A. Supplementary material

Supplementary data associated with this article can be found in the online version of 10.1016/j.cag.2009.10.005.